# Stimulated Magnonic Frequency Combs

*Xueyu Guo[1,#], Tianci Gong[2,#], Guibin Lan[3,#], Mengying Guo[1], Xiufeng Han[3], Guoqiang Yu[3,\*], Peng Yan[2,\*], Qi Wang[1,\*]*

[1] *School of Physics, Hubei Key Laboratory of Gravitation and Quantum Physics, Institute for Quantum Science and Engineering, Huazhong University of Science and Technology, Wuhan, China*

[2] *School of Electronic Science and Engineering and State Key Laboratory of Electronic Thin Films and Integrated Devices, University of Electronic Science and Technology of China, Chengdu, China*

[3] *Beijing National Laboratory for Condensed Matter Physics, Institute of Physics, Chinese Academy of Sciences, Beijing, China.*

**Abstract:**

Magnonic frequency combs, characterized by a series of discrete frequency lines, have emerged as a promising frontier in magnon spintronics, with potential applications in advanced information processing and sensing technologies. Although the three-magnon scattering process is widely recognized as a fundamental mechanism for generating these combs, its experimental realization has remained challenging due to the high threshold power and strict conservation of momentum and energy. In this work, we propose a novel mechanism for the stimulated generation of magnonic frequency combs that overcomes these limitations. Our approach offers precise and efficient control over key comb properties, including spacing between spectral lines and the number of lines, marking a significant advancement in the field. We substantiate this mechanism through a robust combination of theoretical modeling, micromagnetic simulations, and experimental validation. This study not only demonstrates the feasibility of our method but also opens new pathways for integrating magnonic frequency combs into practical spintronic devices.

# These authors contributed equally.
\* Corresponding authors. Email: guoqiangyu@iphy.ac.cn, yan@uestc.edu.cn, williamqiwang@hust.edu.cn

The optical frequency comb represents a groundbreaking advancement in optics, functioning as an ultra-precise ruler for measuring optical frequencies. By generating a perfectly stable and evenly spaced array of reference frequencies, it has enabled transformative applications across multiple fields, including ultra-high-precision spectroscopy, next-generation optical atomic clocks, advanced laser ranging systems, molecular fingerprinting for chemical analysis, and high-bandwidth telecommunications [1-3].

Building on the success of optical frequency combs, spin waves - quantized collective excitations (magnons) of magnetic moments in ordered materials - have recently emerged as a promising platform for frequency comb generation [4-8]. These excitations exhibit intrinsic nonlinearity through three- and four-magnon processes [9-12], nonlinear frequency shifts [13-16], and robust soliton formation [17-19], which are essential for developing coherent frequency combs [12, 20-23]. Compared to optical systems, spin waves offer distinct advantages, including strong nonlinear interactions at microwave frequencies [24-26], high tunability [27,28], and low damping losses [29-31]. These characteristics make spin waves an ideal medium for realizing magnonic frequency combs (MFCs), a novel class of coherent spectra with potential applications in on-chip signal processing, neuromorphic computing, and high-precision magnetometry.

Recent theoretical studies have uncovered an appealing mechanism for generating coherent MFCs. This mechanism leverages the interaction between propagating spin waves and topological magnetic textures - such as skyrmions [32-34], vortices [35], and domain walls [36] - to strongly enhance three-magnon scattering processes [20,37]. Experimental validation in ferromagnetic thin films demonstrated that phase-locked spin-wave dynamics, driven at MHz frequencies, can produce harmonic combs extending over six octaves, reaching up to the 60th harmonic [20]. The coherence of these combs was rigorously confirmed using quantum sensing with nitrogen-vacancy centers in diamond, which resolved phase-stable high-order harmonics and their Rabi oscillations [37]. These results not only demonstrate the feasibility of generating MFCs but also highlight their potential for advancing quantum magnonics and hybrid magnon-photonics systems. High harmonics can be viewed as a special class of MFC with the comb spacing being the driving frequency itself. However, a direct observation of three-magnon induced MFC with tunable comb spacing is still challenging [38], due to its high threshold power and strict conservation of momentum and energy in the thin film.

On the other hand, it has been known that a stimulated three-wave mixing process utilizing a modulation signal can significantly lower the driving power [39-41].

In this Letter, we report the experimental observation of MFC via stimulated three-magnon scattering processes. Compared to conventional three-magnon scattering, stimulated three-magnon scattering is achieved by introducing an additional external drive, significantly enhancing scattering efficiency. In this experiment, a tunable low-frequency modulation signal ($f_m$) is applied to emulate such an external drive. In practical applications, this drive could originate from various environmental sources, serving as a trigger for three-magnon scattering and enabling the sensor's functionality.

In our experiments, we demonstrate the coherent excitation of equispaced spectral lines by superimposing a tunable low-frequency modulation signal ($f_m$) onto an excitation wave ($f_e$) in a nonlinear magnonic system. Notably, this method provides two independent control dimensions: the comb spacing ($\Delta f$), which is precisely locked to the modulation frequency ($\Delta f = f_m$), and the number of spectral lines, which scales with the modulation power. These features enable precise, lower power, real-time control over the MFC spectra, making them adaptable for a wide range of applications.

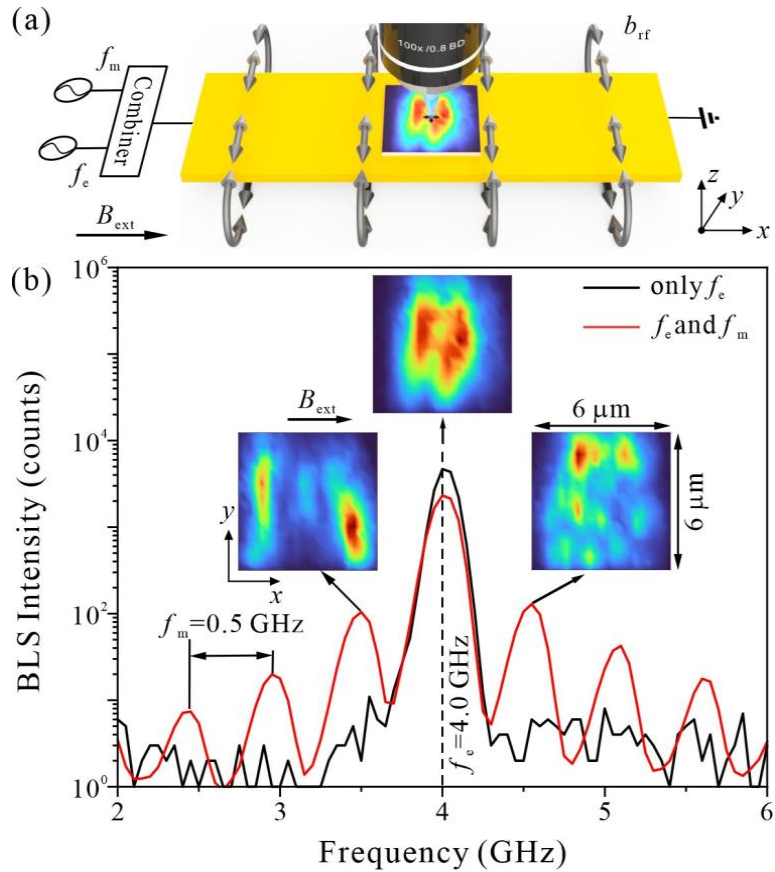

*Figure 1. (a) **Schematic of the experimental configuration.** A 5 μm×5 μm×15 nm NiFe film is positioned at the center of a gold microwave antenna (yellow) on a silicon substrate. Two synchronized microwave sources generate signals at the excitation frequency ($f_e$) and modulation frequency ($f_m$), combined through a power combiner. An in-plane bias field ($B_{ext}$) is applied along the antenna's longitudinal axis (x-axis), while the microwave Oersted field ($b_{rf}$) oscillates along the y-axis to excite magnons. A microfocused Brillouin light scattering (μBLS) setup detects the spatially-resolved spin-wave spectra. (b) Comparison of measured spectra. The black curve represents the spectrum under single-frequency excitation at $f_e$, while the red curve illustrates the stimulated magnonic frequency comb generated by dual-frequency excitation ($f_e + f_m$). Insets depict the spatial intensity distributions of three key frequency components: (i) the direct excitation mode at $f_e$, (ii) the sum-frequency mode at $f_e + f_m$, and (iii) the difference-frequency mode at $f_e - f_m$.*

Figure 1(a) illustrates the experimental setup, featuring a 5 μm×5 μm×15 nm NiFe (Permalloy) ferromagnetic element lithographically patterned at the center of a gold microwave antenna. Two synchronized microwave sources generate the excitation ($f_e$) and modulation ($f_m$) signals, which are combined using a broadband power combiner for independent frequency and amplitude control. An in-plane bias field ($B_{ext}$) is applied along the antenna's longitudinal ($x$) axis, while the microwave-driven Oersted field ($b_{rf}$) oscillates along the $y$-axis to excite magnons. Microfocused Brillouin light scattering (μBLS) spectroscopy measures the resulting magnetization dynamics with high frequency and spatial resolution.

Figure 1(b) displays the BLS spectra, averaged across the entire Permalloy structure, for two distinct excitation conditions under an external field $B_{ext}$=18 mT: single-frequency (black curve, $f_e$=4.0 GHz, $P_e$=20 dBm) and dual-frequency excitation (red curve, $f_e$=4.0 GHz, $P_e$=20 dBm with an additional modulation signal at $f_m$=0.5 GHz, $P_m$=30 dBm). The excitation frequency of 4.0 GHz was deliberately chosen because it closely aligns with the ferromagnetic resonance (FMR) frequency under these experimental conditions, optimizing excitation efficiency. For additional insight into the dispersion characteristics, the corresponding simulated and analytical dispersion curve of the Permalloy square is available in the Supplementary Materials [42]. In the single-frequency spectrum (black curve), a solitary resonance peak appears at 4.0 GHz, consistent with the applied excitation frequency. By contrast, the dual-frequency excitation (red curve) produces a striking MFC—a series of equidistant spectral peaks with a spacing of $\Delta f = f_m = 0.5$ GHz. Most notably, this comb includes frequency components below the FMR frequency, such as the difference frequency $(f_m - f_e) = 3.5$ GHz, which lies in a region typically inaccessible under conventional excitation methods. This unexpected feature highlights the dual-frequency approach's ability to probe and manipulate magnon states beyond traditional limits.

To investigate the spatial properties of this comb, we scanned a laser spot with a diameter of approximately 350 nm across the entire Permalloy square, with a step size of 250 nm. We extracted two-dimensional spin-wave intensity distributions for three key spectral components: the direct excitation frequency ($f_e$), the sum frequency ($f_e + f_m$), and the difference frequency ($f_e - f_m$). The resulting 2D spatial maps, presented as insets in Fig. 1(b), reveal distinct localization patterns. As expected, the resonance for the direct excitation frequency (4.0 GHz) is concentrated at the center of the square, corresponding to a quasi-ferromagnetic resonance with wavenumber $k \approx 0$. In contrast, the difference-frequency resonance (3.5 GHz) is localized along the edges perpendicular to the external field direction, indicating the edge-localized spin-wave modes. These edge modes arise from internal field potential well created by the demagnetizing field, which lowers the resonance frequency below the FMR frequency [43]. The sum frequency resonance (4.5 GHz) exhibits standing wave behavior along both the $x$- and $y$-axes ($k_x \neq 0$, $k_y \neq 0$), a consequence of the square structure's boundary conditions. Notably, the spatial distributions of all observed peaks in the comb share characteristics with these primary modes, further supported by micromagnetic simulations in the Supplementary Materials [42].

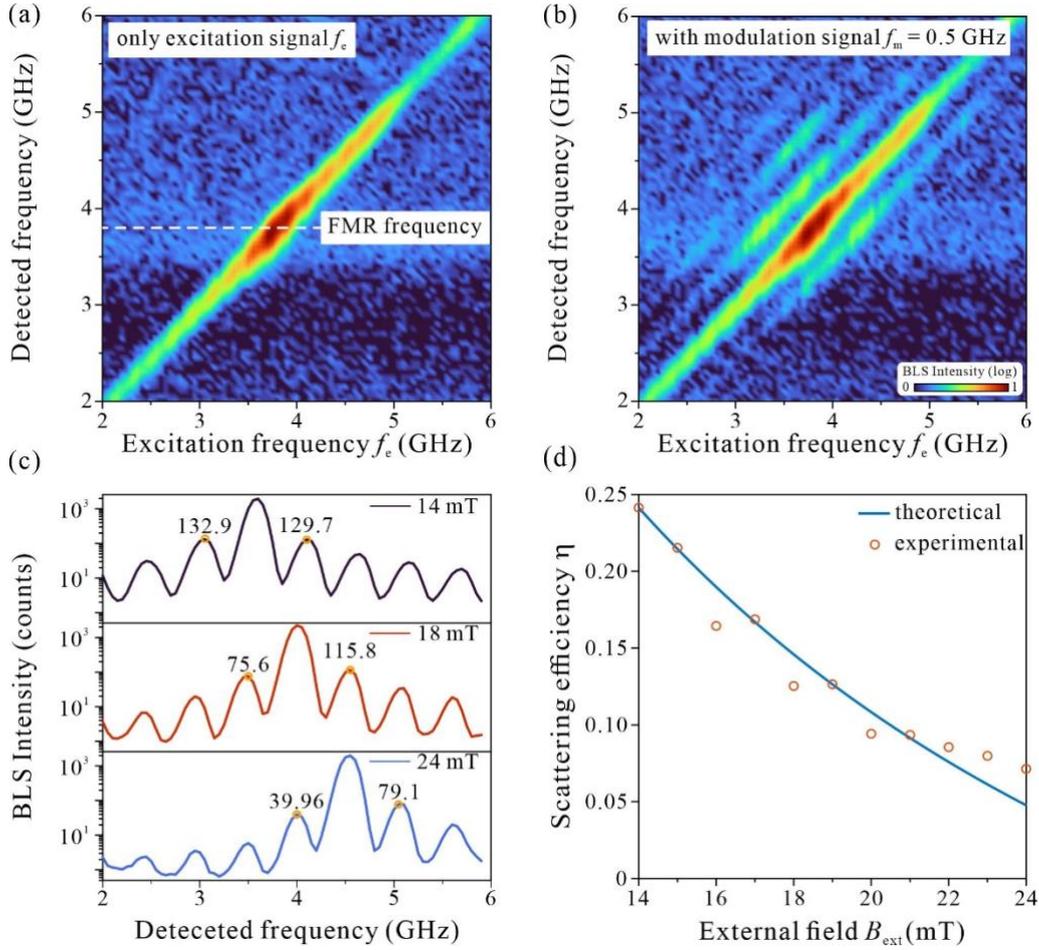

*Figure 2. (a) Normalized two-dimensional (2D) frequency-domain distribution obtained from a frequency-swept excitation signal (2.0–6.0 GHz, 20 dBm) applied without modulation. (b) Corresponding 2D frequency-domain distribution when the excitation signal (2.0–6.0 GHz sweep) is combined with a modulation signal ($f_m$=0.5 GHz, 30 dBm). The comparison between (a) and (b) highlights the critical role of the modulation signal in facilitating the generation of the stimulated MFC. (c) The BLS spectra measured under varying external magnetic field. (d) Theoretical (bule curve) and experimental (red circles) results for the scattering efficiency $\eta = (|a_p| + |a_q|)/|a_e|$ as a function of the external magnetic field.*

As shown in Fig. 1(b), the experimental results demonstrate that a modulation frequency of $f_m$=0.5 GHz effectively triggers nonlinear magnonic phenomena, leading to the formation of an MFC. To explore the comb generation mechanism, we conducted comparative experiments across a wide excitation frequency range (2-6 GHz) with and without the modulation, under a constant external field of $B_{ext}$ =18 mT. Figure 2(a) presents the BLS-detected frequency as a function of the excitation frequency, revealing a single linear response where the input and detected frequencies correspond one-to-one. The horizontal dashed line marks the quasi-FMR frequency at approximately 3.8 GHz under this field. While excited and thermal signals exhibit weak response below this quasi-FMR frequency, a pronounced excitation peak emerges precisely at 3.8 GHz, confirming that efficient excitation occurs predominantly at the FMR frequency. Figure 2(b) shows the spectral response when both excitation and modulation frequencies ($f_m$=0.5 GHz, $P_m$=30 dBm) are applied simultaneously. Multiple evenly spaced spectral lines appear, with their spacing matching the modulation frequency $f_m$, unequivocally confirming the MFC's formation. This demonstrates precise control over the comb spacing by tuning $f_m$, underscoring the potential for tailored magnonic spectral engineering. Similar behavior is observed for modulation frequencies ranging from 0.4 to 0.6 GHz as detailed in the Supplementary Materials [42]. It is important to note that the modulation frequency used here is significantly lower than the FMR frequency, distinguishing this approach from four-magnon scattering-induced MFCs, which require modulation frequencies exceeding the FMR frequency [12]. Micromagnetic simulations, presented in the Supplementary Materials [42], further validate this modulation-induced comb structure. The underlying mechanism is attributed to stimulated three-magnon scattering processes, where nonlinear magnon interactions drive the energy redistribution among spectral components.

In order to describe the processes, we consider a single-frequency driving system with the Hamiltonian

$$H_0 = \sum_k \omega_k a_k^\dagger a_k + h_1 (a_k e^{i\omega_m t} + a_k^\dagger e^{-i\omega_m t}), \tag{1}$$

where $a_k(a_k^\dagger)$ is the magnon annihilation (creation) operator, while $h_1$ and $\omega_m = 2\pi f_m$ ($<\omega_k$) represent the amplitude and angular frequency of the modulations signal, respectively. The eigen-frequency ($\omega_k$) of magnon reads [44,45]

$$\omega_k = \sqrt{(\gamma H + A^* k^2)(\gamma H + A^* k^2 + \gamma \mu_0 M_s F_p)}, \tag{2}$$

where $A^* = 2\gamma A/M_s$ in terms of gyromagnetic ratio $\gamma$, exchange constant $A$, and saturated magnetization $M_s$. We also use the dipole-dipole matrix element $F_p = 1 - g_k \cos^2\theta_k + \gamma \mu_0 M_s g_k (1-$

$g_k)\sin^2\theta_k/(\gamma H+A^*k^2)$ with $g_k=1-[1-e^{-kd}]/(kd)$, thickness $d$ of the film, external field $H$, and angle $\theta_k$ between the spin-wave wavevector $k$ and the magnetization direction $\mathbf{M}$.

The global driving field selectively couples to the ferromagnetic resonance (FMR) mode $\omega_f = \omega_k(k=0)$, with the Hamiltonian

$$H_0=\omega_f a_m^\dagger a_m+h_1(a_m e^{i\omega_m t}+a_m^\dagger e^{-i\omega_m t}). \tag{3}$$

Since the modulation frequency lies below the lowest eigen-frequency (FMR frequency $\omega_f$), the thin-film response primarily localizes at the boundaries and decays toward the center.

Based on the Hamiltonian (3), we readily derive the Heisenberg dynamic equation for the system as

$$i\frac{da_m}{dt}=(\omega_f-i\alpha_f\omega_f)a_m+h_1 e^{-i\omega_m t}. \tag{4}$$

Here, $\alpha_f$ are the effective Gilbert damping constants of the corresponding magnon mode. With $a_m=|a_m|e^{i\omega_m t}$, we observe the amplitude of the stimulated magnon mode

$$|a_m|=\frac{h_1}{\sqrt{(\omega_f-\omega_m)^2+\alpha_f^2\omega_f^2}}. \tag{5}$$

The stimulated magnon amplitude is suppressed when $\omega_m$ deviates further from $\omega_f$. To enhance three-magnon processes, we apply a second driving single (central single) at $\omega_e = 2\pi f_e$, with the corresponding Hamiltonian [32,33]

$$H=\omega_e(a_e^\dagger a_e+a_m^\dagger a_m+a_q^\dagger a_q)+\omega_p a_p^\dagger a_p+g_p(a_e a_m a_p^\dagger+H.c.)+g_q(a_e a_m^\dagger a_q^\dagger+H.c.)$$
$$+h_1(a_m e^{i\omega_m t}+a_m^\dagger e^{-i\omega_m t})+h_2(a_e e^{i\omega_e t}+a_e^\dagger e^{-i\omega_e t}), \tag{6}$$

where $g_p$ and $g_q$ are the coupling strength of the three-magnon confluence and splitting, respectively. Under dual frequency driving, the linear response of system excites two distinct modes ($\omega_e$ and $\omega_m$), while nonlinear effects would mix two modes with each other and generate the confluence mode (sum-frequency: $\omega_p = \omega_e + \omega_m$) and splitting mode (difference-frequency: $\omega_q = \omega_e - \omega_m$). These two secondary signals further hybridize with $\omega_m$ to generate higher-order frequency modes, eventually leading to the MFC.

Figure 2(c) presents spectra obtained under varying external magnetic fields which correspond to different FMR frequencies. These spectra illustrate the intensity evolution of both the confluence and splitting modes - key features of the MFC. The curves were generated by averaging the signal across the entire Permalloy structure, resulting in higher side-mode intensities compared to the center-focused 2D BLS spectra in Fig. 2(b). As indicated by the orange circles, the intensities of these modes decrease with increasing external field strength. This behavior can be quantitatively understood by solving the coupled Heisenberg dynamic equations, which model the nonlinear interactions governing the mode amplitudes.

$$i\frac{da_e}{dt} = (\omega_e - i\alpha_e\omega_e)a_e + g_p a_m^\dagger a_p + g_q a_m a_q + h_2 e^{-i\omega_e t},$$

$$i\frac{da_m}{dt} = (\omega_f - i\alpha_f\omega_f)a_m + g_p a_e^\dagger a_p + g_q a_e a_q^\dagger + h_1 e^{-i\omega_m t},$$

$$i\frac{da_p}{dt} = (\omega_p - i\alpha_p\omega_p)a_p + g_p a_e a_m,$$

$$i\frac{da_q}{dt} = (\omega_f - i\alpha_f\omega_f)a_q + g_q a_e a_m^\dagger. \tag{7}$$

where we assume that all magnon modes have identical damping rates $\alpha_e = \alpha_f = \alpha_p = \alpha = 0.007$. Since the frequency of the splitting mode $\omega_q$ lies below the FMR frequency $\omega_f$, the splitting mode ($\omega_q$) retains partial characteristics of stimulated vibration, exhibiting a faster decrease in amplitude with the external field compared to the confluence mode ($\omega_p$) [see Fig. 2(c)]. When consider a steady-state solution with $\frac{da_v}{dt} = i\omega_v \langle a_v \rangle e^{-i\omega_v t}$ ($v$=e,m,p,q), one can obtain

$$\langle a_p \rangle = \frac{g_p \langle a_e \rangle \langle a_m \rangle}{i\alpha_p \omega_p},$$

$$\langle a_q \rangle = -\frac{g_q \langle a_e \rangle \langle a_m^\dagger \rangle}{(\omega_f + \omega_m - \omega_e - i\alpha_f \omega_f)}, \quad (8)$$

while $\langle a_e \rangle$ and $\langle a_m \rangle$ are fully determined by the following self-consistent equations,

$$0 = (-i\alpha_e \omega_e)\langle a_e \rangle + \langle a_e \rangle |\langle a_m \rangle|^2 \left(g_p^2 \frac{1}{i\alpha_p \omega_p} - g_q^2 \frac{1}{(\omega_f + \omega_m - \omega_e - i\alpha_f \omega_f)}\right) + h_2,$$

$$0 = (\omega_f - \omega_m - i\alpha_m \omega_f)\langle a_m \rangle + \langle a_m \rangle |\langle a_e \rangle|^2 \left(g_p^2 \frac{1}{i\alpha_p \omega_p} - g_q^2 \frac{1}{(\omega_f + \omega_m - \omega_e + i\alpha_f \omega_f)}\right) + h_1 \quad (9)$$

Through fitting of simulation data, we determine the coupling strengths $g_p = g_q = 0.02$ GHz, enabling numerical solutions for the dependence of the amplitude on the system parameters. Here, we defined a scattering efficiency as $\eta = (|\langle a_p \rangle| + |\langle a_q \rangle|)/|\langle a_e \rangle|$. As shown by the theoretical curve in Fig. 2(d), the scattering efficiency $\eta$ decreases monotonically with increasing external field (FMR frequency), consistent with experimental observations. This behavior originates from the amplitude attenuation of the modulation signal at higher FMR frequencies, as quantitatively described by Eq. (5). The resulting suppression of the modulation amplitude consequently diminishes the nonlinear mixing efficiency between the modulation signal and the central mode, thereby reducing the intensity of both confluence and splitting modes.

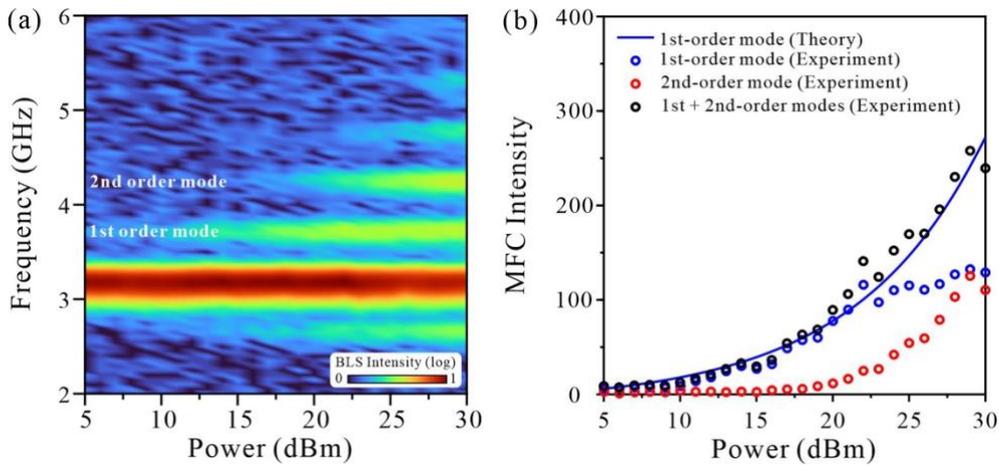

*Figure 3 **Power-dependent characteristics of the stimulated MFC.** (a) Evolution of the comb spectrum as the modulation signal power increases. The excitation signal is fixed at 3.2 GHz with a power of 15 dBm, while the modulation signal, set at 0.5 GHz, has its power swept from 5 to 30 dBm. (b) Comparison of*

*theoretical (solid blue curve) and experimental results of the first-order mode (blue circles), second-order mode (red circles), and their combined intensity (black circles) as a function of modulation power.*

By numerically solving Eq. (7), the intensities of both the confluence and splitting modes are shown to exhibit a power-dependent relationship with the modulation signal strength. Figure 3 presents the BLS measurements of frequency intensity versus the power of modulation frequency, demonstrating the continuous evolution of the stimulated MFC. The measurements were performed with a fixed 3.2 GHz (corresponding to the FMR of external field of 14 mT) excitation signal (15 dBm) while sweeping the power of 0.5 GHz modulation signal from 5 to 30 dBm. The results reveal a smooth, monotonic increase in both the number and intensity of comb lines with increasing modulation power throughout the entire tested range.

To quantitatively characterize the power dependence, we extract the intensity of the first-order frequency component ($f_e+f_m$) by integrating the spectral signal within the frequency range delimited by the two vertical dashed lines in Fig. 3(a). These integrated intensities, plotted as blue circles in Fig. 3(b), reveal a clear power-dependent evolution of the mode. The experimental data reveal a distinct two-stage behavior: at low modulation powers, the first-order intensity increases monotonically with power, in excellent agreement with theoretical predictions [blue curve in Fig. 3(b)]. However, at high powers, the experimental intensity saturates, while the theoretical curve continues to rise, highlighting a significant discrepancy. This deviation arises because the theoretical model does not account for higher-order nonlinear effects, specifically, the energy transfer from the first-order mode ($f_e+f_m$) to the second-order mode ($f_e+2f_m$) which becomes prominent at elevated powers.

To confirm this interpretation, we extract and plot intensity of the second-order mode ($f_e+2f_m$), shown as red circles in Fig. 3(b). The data reveal a clear power correlation: the onset of the second-order mode aligns precisely with the saturation of the first-order mode, providing direct evidence of energy transfer between them. This energy transfer drives the formation of a MFC through a cascaded nonlinear process. Specifically, the primary magnon excitation at $f_e$ interacts with the modulation signal at $f_m$ to produces first-order sidebands at $f_e±f_m$, which then successively interact with $f_m$ to generate higher-order comb teeth, consistent with theoretical predictions. Furthermore, when the combined intensity of the first- and second-order modes is analyzed [black circles in Fig. 3(b)], the results closely match the theoretical model, reinforcing this interpretation. Notably, the spectral range of the nonlinear comb generation follows a power-dependent scaling law, where changes in modulation power alter the number of comb teeth in the MFC, consequently modifying the frequency range covered by the teeth. Thus, precise tuning of the modulation power enables deterministic control over the number of resolvable comb teeth.

In summary, this work presents a novel approach for generating a tunable magnonic frequency comb via stimulated three-magnon scattering. We develop an analytical model to elucidate the underlying mechanism, rigorously validated by experimental results and micromagnetic simulations. Unlike previous techniques, our approach allows the modulation frequency to be set either above or below the ferromagnetic resonance frequency, significantly enhancing operational flexibility and broadening the scope of MFC applications. Furthermore, our method provides precise and efficient control over the comb's spectral properties: the tooth spacing is directly tuned by the modulation frequency, while the number of teeth is optimized through modulation power adjustments. This advancement represents a pivotal step toward the practical implementation of MFC in next-generation magnonic sensors and spintronic devices.

**Acknowledgments**

This work was supported from the National Key Research and Development Program of China (Grant No. 2023YFA1406600 and No. 2022YFA1402802) and the National Natural Science Foundation of China (NSFC) (Grant No. 12574118 and No.12374103).

**References**

1. S. A. Diddams, K. Vahala, and T. Udem, Optical frequency combs: Coherently uniting the electromagnetic spectrum, *Science.* **369**, 267 (2020).
2. S. A. Diddams, L. Hollberg, and V. Mbele, Molecular fingerprinting with the resolved modes of a femtosecond laser frequency comb., *Nature* **445**, 627 (2007).
3. L. Chang, S. Liu, and J. E. Bowers, Integrated optical frequency comb technologies, *Nat. Photon.* **16**, 95 (2022).
4. A. V. Chumak, P. Kabos, M. Wu, C. Abert, C. Adelmann, et al., Advances in Magnetics Roadmap on Spin-Wave Computing, *IEEE Trans. Magn.* **58**, 1 (2022).

5. X. Han, H. Wu, T. Zhang, Magnonics: Materials, physics, and devices, *Appl. Phys. Lett.* **125**, 020501 (2024).

6. Q. Wang, G. Csaba, R. Verba, A. V. Chumak, and P. Pirro, Nanoscale magnonic networks, *Phys. Rev. Applied* **21**, 040503 (2024).

7. P. Pirro, V. I. Vasyuchka, A. A. Serga, and B. Hillebrands, Advances in coherent magnonics, *Nat. Rev. Mater.* **6**, 1114 (2021).

8. G. Csaba, Á. Papp, and W. Porod, Perspectives of Using Spin Waves for Computing and Signal Processing, *Phys. Lett. A* **381**, 1471 (2017).

9. R. E. Camley, Three-magnon processes in magnetic nanoelements: Quantization and localized mode effects, *Phys. Rev. B* **89**, 214402 (2014).

10. M. Mohseni, Q. Wang, B. Heinz, M. Kewenig, M. Schneider, et al., Controlling the nonlinear relaxation of quantized propagating magnons in nanodevices, *Phys. Rev. Lett.* **126**, 097202 (2021).

11. L. Körber, C. Heins, I. Soldatov, R. Schäfer, A. Kákay, et al., Modification of three-magnon splitting in a flexed magnetic vortex, *Appl. Phys. Lett.* 122, 092401 (2023).

12. T. Hula, K. Schultheiss, F. J. T. Gonçalves, L. Körber, M. Bejarano, et al., Spin-wave frequency combs, *Appl. Phys. Lett.* 121, 112404 (2022).

13. Q. Wang, M. Kewenig, M. Schneider, R. Verba, F. Kohl, et al., A magnonic directional coupler for integrated magnonic half-adders, *Nat. Electron.* **3**, 765 (2020).

14. S. Zheng, Z. Wang, Y. Wang, F. Sun, Q. He, et al., Tutorial: Nonlinear magnonics, *J. Appl. Phys.* **134**, 151101 (2023).

15. Qi Wang, Roman Verba, Björn Heinz, Michael Schneider, Ondřej Wojewoda, Deeply nonlinear excitation of self-normalized short spin waves, *Sci. Adv.* **9**, eadg4609 (2023).

16. H. Merbouche, H. Merbouche, B. Divinskiy, K. O. Nikolaev, C. Kaspar, W. H. P. Pernice, et al., Giant Nonlinear Self-Phase Modulation of Large-Amplitude Spin Waves in Microscopic YIG Waveguides, *Sci. Rep.* **12**, 7246 (2022).

17. A. M. Kosevich, B. A. Ivanov, A. S. Kovalev, Magnetic solitons, *Phys. Rep.* **194**, 117 (1990).

18. M. Bauer, O. Büttner, S. Demokritov, B. Hillebrands, V. Grimalsky, et al., Observation of spatiotemporal self-focusing of spin waves in magnetic films, *Phys. Rev. Lett.* **81**, 3769 (1998).

19. M. Elyasi, K. Sato, and G. E. W. Bauer, Topologically nontrivial magnonic solitons, *Phys. Rev. B* **99**, 13442 (2019).

20. C. Koerner, R. Dreyer, M. Wagener, N. Liebing, H. G. Bauer et al., Frequency multiplication by collective nanoscale spin-wave dynamics, *Science* **375**, 1165 (2022).

21. C. Wang, J. Rao, Z. Chen, K. Zhao, L. Sun, et al., Enhancement of magnonic frequency combs by exceptional points, *Nat. Phys.* **20**, 1139 (2024).

22. X. Liang, Y. Cao, P. Yan, and Y. Zhou, Asymmetric magnon frequency comb, *Nano Lett.* **24**, 6730 (2024).

23. Z. Liu, J. Peng, and H. Xiong, Generation of magnonic frequency combs via a two-tone microwave drive, *Phys. Rev. A* **107**, 053708 (2023).

24. V. E. Demidov, J. Jersch, K. Rott, P. Krzysteczko, G. Reiss, et al., Nonlinear Propagation of Spin Waves in Microscopic Magnetic Stripes, *Phys. Rev. Lett.* **102**, 177207 (2009).

25. V. S. L'vov, *Wave Turbulence under Parametric Excitation* (Springer Nature, 1994).

26. K. O. Nikolaev, S. R. Lake, G. Schmidt, S. O. Demokritov, and V. E. Demidov, Resonant generation of propagating second-harmonic spin waves in nano-waveguides, *Nat. Commun.* **15**, 1827 (2024).

27. A. V. Chumak, A. A. Serga, and B. Hillebrands, Magnon Transistor for All-Magnon Data Processing, *Nat. Commun.* **5**, 4700 (2014).

28. R. Verba, M. Carpentieri, G. Finocchio, V. Tiberkevich, and A. Slavin, Amplification and Stabilization of Large-Amplitude Propagating Spin Waves by Parametric Pumping, *Appl. Phys. Lett.* **112**, 042402 (2018).

29. H. Qin, R. B. Holländer, Lukáš Flajšman, F. Hermann, R. Dreyer, et al., Nanoscale Magnonic Fabry-Pérot Resonator for Low-Loss Spin-Wave Manipulation, *Nat Commun.* **12**, 2293 (2021).

30. C. Dubs, O. Surzhenko, R. Linke, A. Danilewsky, U. Brückner, et al., Sub-micrometer yttrium iron garnet LPE films with low ferromagnetic resonance losses, *J. Phys. D: Appl. Phys.* **50**, 204005 (2017).

31. C. Liu, J. Chen, T. Liu, F. Heimbach, H. Yu, et al., Long-distance propagation of short-wavelength spin waves, *Nat. Commun.* **9**, 738 (2018).

32. X. Yao, Z. Jin, Z. Wang, Z. Zeng, and P. Yan, Terahertz magnon frequency comb, *Phys. Rev. B* **108**, 134427 (2023).

33. Z. Wang, H. Y. Yuan, Y. Cao, Z.-X. Li, R. A. Duine, et al., Magnonic Frequency Comb through Nonlinear Magnon-Skyrmion Scattering, *Phys. Rev. Lett.* **127**, 037202 (2021).

34. Y. Liu, T. Liu, Q. Yang, G. Tian, Z. Hou, et al., Design of controllable magnon frequency comb

in synthetic ferrimagnets, *Phys. Rev. B*. **109**, 174412 (2024).

35. Z. Wang, H. Y. Yuan, Y. Cao, and P. Yan, Twisted Magnon Frequency Comb and Penrose Superradiance, *Phys. Rev. Lett.* **129**, 107203 (2022).

36. Z. Zhou, X. Wang, Y. Nie, Q. Xia, and G. Guo, Spin wave frequency comb generated through interaction between propagating spin wave and oscillating domain wall, *J. Magn. Magn. Mater.* **534**, 168046 (2021).

37. G. Lan, K.-Y. Liu, Z. Wang, F. Xia, H. Xu, et al., Coherent harmonic generation of magnons in spin textures, *Nat Commun.* **16**, 1178 (2025).

38. C. Heins, A. Kákay, J.-V. Kim, G. Hlawacek, J. Fassbender, et al, Control of Magnon Frequency Combs in Magnetic Rings, arXiv:2501.05080.

39. A. V. Gaponov, Instability of a system of excited oscillators with respect to electromagnetic perturbations, *Sov. Phys. JETP* **12**, 232 (1961)

40. V. M. Fain, A quantum generalization of the expression for energy dissipation, *Sov. Phys. JETP* **23**, 882 (1966)

41. B. Fain and P. W. Milonni, Classical stimulated emission, *J. Opt. Soc. Am. B* **4**, 78 (1987)

42. Supplemental Material

43. V. E. Demidov, J. Jersch, and S. O. Demokritov, Transformation of propagating spin-wave modes in microscopic waveguides with variable width, *Phys. Rev. B* **79**, 054417 (2009).

44. B. A. Kalinikos and A. N. Slavin, Theory of dipole-exchange spin wave spectrum for ferromagnetic films with mixed exchange boundary conditions, *J. Phys. C: Solid State Phys.* **19**, 7013 (1986).

45. T. Brächer, P. Pirro, and B. Hillebrands, Parallel pumping for magnon spintronics: Amplification and manipulation of magnon spin currents on the micron-scale, *Phys. Rep.* **699**, 1 (2017).